\documentclass[12pt]{article}
\usepackage{latexsym}
\usepackage{amsmath}
\usepackage{amsfonts}
\usepackage{amssymb}
\usepackage[cp1250]{inputenc}

\title{Euclidean Path Integral and Higher-Derivative Theories}
\author{Krzysztof Andrzejewski\thanks{supported by the grant 690 and 1037 of the University of {\L}\'od\'z.}, 
 Joanna Gonera\thanks{supported by the grant  689 and 1037 of the University of {\L}\'od\'z.}, 
Pawe{\l}  Ma\'slanka\thanks{supported by the grant 690 and 1037 of the University of {\L}\'od\'z.}\\ 
Department of Theoretical Physics II \\
University of {\L}\'od\'z \\
Pomorska 149/153, 90 - 236 {\L}\'od\'z, Poland.}

\date{}

\begin{document}
\maketitle
\begin{abstract}
We consider the Euclidean path integral approach to higher-derivative theories proposed by Hawking and Hertog
( Phys. Rev. \underline{D65} (2002) 103515). The Pais-Uhlenbeck oscillator is studied in some detail. The operator
algebra is reconstructed and  the structure of the space of  states revealed. It is shown that the quantum theory results from quantizing the classical complex dynamics in which the original dynamics is consistently immersed. The field-theoretical counterpart of Pais-Uhlenbeck 
oscillator is also considered.

\end{abstract}

\newpage

\section {Introduction}

Theories described by Lagrangians containing higher derivatives seem to be of some relevance to physics. An interesting example is
provided by Einstein gravity supplied by the terms containing higher powers of curvature \cite{b1}. Other examples include string theory with
 extrinsic curvature \cite{b2},\cite{b3}, anyons \cite{b3}, field theory on noncommutative space-time  \cite{b4} or theory of
 radiation reaction \cite{b5}.\\
\hspace*{15pt} The question whether a viable quantum counterpart of higher-derivative theory exists provides a real challenge. The straightforward
canonical quantization based on Ostrogradski Hamiltonian formalism \cite{b6} suffers from serious drawback: the energy is unbounded from below.
If one tries to manage  this problem the ghost states seem to enter inevitably the theory; this is thought to be a fatal flaw (cf. \cite{b7})
 unless
it can be cured due to the existence of specific symmetry restoring unitarity in physical sector.\\

It is sometimes claimed that the energy does not need to be bounded from below. In fact one can construct models which are stable in spite of
 the fact that their energy is unbounded from below. However, one should keep in mind that our models provide usually some idealization which 
consists in neglecting various perturbative terms that could easily spoil stability. On the other hand, there exists at least one general and 
physically relevant example of class of theories which contain ghosts but the unitarity in physical subspace is preserved for any physically 
acceptable perturbation; these are gauge theories. \\

In general, in order to deal properly with the  quantization problem one has to have a clear understanding of what is meant by correct quantization 
of a given classical theory. There are two natural conditions to be fulfilled: 

$(i)$\ the resulting structure posses all relevant properties of quantum
theory ( superposition principle, reality of admissible values of measurable quantities, probabilities ranging between $0$\ and $1$, etc.);

${(ii)}$\ in the limit $\hslash \rightarrow 0$\ the initial classical theory should be recovered; in particular, the expectation values of basic
 variables  for properly selected states (coherent states) obey in this limit the classical dynamical equations. 

The latter condition implies that 
the quantum  theory is not only defined by the equations of motion and commutation rules but also by the transformation properties of dynamical 
variables under hermitean ( with respect to the relevant scalar product ) conjugation. 
This fact should be properly recognized when considering the admissible quantization schemes. It is often stressed that the
 Hamiltonian is not only defined by the formal differential operator but also by the boundary conditions imposed; this is fairly obvious.
However, it is equally obvious that the latter cannot be imposed at will. In fact, the choice of boundary conditions influences the spectral
properties of other dynamical variables (in particular - coordinates) which should in turn be chosen in accordance with the properties of their
classical counterparts. For example, imposing in the case of harmonic oscillator the integrability condition along imaginary axis one obtains 
purely discrete negative spectrum. However, this is not a proper way to quantize the classical harmonic oscillator as it contradicts the reality
of coordinate variable (which is inherent in definition of harmonic oscillator). \\

The celebrated Pais-Uhlenbeck (PU) oscillator \cite{b8}
 provides an even more distinct example.
 In the oscillatory regime it can be written ( within Ostrogradski framework ) as a difference of two harmonic oscillators. 
Imposing the purely imaginary boundary condition 
on the second oscillator one finds the energy of PU system positive. However, the original coordinate is a linear combination of oscillators
 coordinates and hence takes in general arbitrary complex values. On the other hand, the original coordinate of classical PU oscillator is 
 real. Therefore,  the choice of boundary conditions for the Hamiltonian enforcing the expected properties of the energy spectrum 
does not reflect the properties of other dynamical variables.
 If this is the case
we are quantizing the classical theory which differs from the one we have started with;  changing the initial theory is not a proper remedy 
for the problems we encounter.\\

One can pose the question to what extent is the quantization procedure  flexible; for example, whether some kind of complexification is admissible.
For standard Lagrangians the dynamical equations are of second order in time derivatives. Therefore, classically the state of the system is
determined uniquely once we know $q_i(t)$\ and $\dot q_i(t)$\ at one given moment $t$. Any quantity of physical relevance (usually related to 
some Noether charge)  can be constructed out of $q_i(t)$\ and $\dot q_i(t)$; moreover, the reality of $q_i(t)$\ and $\dot q_i(t)$\ for a given
$t$\ implies the reality of $q_i(t)$\ and all its derivatives for any $t$. Thus $q_i$\ and $\dot q_i$\  should be viewed as basic dynamical
 variables which, at the quantum level, generate the irreducible set of  observables. Now, the problem is that, on the one hand,
 $q_i(t)$\ and $\dot q_i(t)$\ are related by time derivative operation while, on the other, they can attain arbitrary values at a fixed 
initial moment. The way out is well known: it is the Hamiltonian formalism where $\dot q_i(t)$\ is replaced by $p_i$\ which is related to  
$\dot q_i(t)$\ only "on-shell". Generalizing, one can say that the first step toward the quantization is to replace $\dot q_i$\ by another
variable in such a way that: $(a)$\ it is related to  $\dot q_i$\  only "on-shell";  $(b)$\ the real character of $q_i$\ 
and $\dot q_i$\  is preserved even if the additional variable takes complex values.
 
In the case of Lagrangians containing also second order time derivatives the basic variables determining uniquely the classical state 
are $q, \dot q, \ddot q$\ and $ \dddot q $. Therefore, according to the same line of reasoning one
 should immerse the initial dynamical system into a "Hamiltonian" one in such a way that  $q, \dot q, \ddot q, \dddot q $\ are replaced by new 
 variables related to the former ones only "on-shell". The new dynamics may be complex provided $q(t)$\ remains  real for all $t$\ . So there
 is some freedom in the choice of the extended dynamics; the consistency condition, apart from reality of $q(t)$\, is that $q(t)$\ obeys
 the dynamical equation(s) following from the initial action principle. Such an extended dynamics provides the starting point for quantization.\\
 
 The Ostrogradski formalism is the standard way of constructing the extended dynamics. All variables, including the auxiliary ones, are here 
real and the theory is a real Hamiltonian one. Straightforward quantization leads to the Hamiltonian unbounded from below and the space 
of states which is a Hilbert space (no negative norm states). One can also consider complex extensions of original dynamics. Such a possibility
 has been suggested some time ago \cite{b9}, \cite{b10}. It was found \cite{b9} to lead to positive energy and indefinite metric in the space
 of states.
 
 On the other hand, Hawking and Hertog \cite{b11} proposed the quantization method for higher-derivative theories based on Euclidean 
path integral in Lagrangian formulation. Our aim here is to study the Hawking-Hertog method in some detail.We compute the exact 
propagator for Pais-Uhlenbeck model and reconstruct the underlying quantum-mechanical algebra.The structure of the space of states is revealed and
 the existence of ghosts is shown to follow from the structure of classical action. Hamiltonian appears to be nonhermitean with respect to the
 standard scalar product. In spite of that it posses, in the oscillatory regime, purely real positive point spectrum.
 Outside this regime the eigenvalues cease to be real (at the crossover point the Hamiltonian is even nondiagonalizable) but the path integral
continues to represent the propagator; the very existence of propagator is related to the fact that the real parts of energy eigenvalues are 
positive.
We consider also the field-theoretic counterpart of PU oscillator.

\section{The Pais -Uhlenbeck model}

In the Hawking-Hertog approach \cite{b11} one starts with the Lagrangian 
\begin{eqnarray}
L(t)\equiv L\left(q(t),\dot q(t),\ddot q(t)\right)   \label{w1}
\end{eqnarray}
and assumes that its Euclidean counterpart
\begin{eqnarray}
L_E(t)\equiv -L(-it)   \label{w2}
\end{eqnarray}
gives rise to positive-definite action
\begin{eqnarray}
S_E= \int\limits_{0}^{t}d\tau L_E(\tau )  \label{w3}
\end{eqnarray}
According to the suggestion of Hawking and Hertog  one expects the Euclidean path integral $\int Dqe^{-S_E[q]}$\ to converge and to give
rise to  a well-defined Euclidean quantum theory.\\
 Our aim here is to analyse the real-time counterpart of the Hawking-Hertog construction for the Pais-Uhlenbeck Lagrangian:
\begin{eqnarray}
L= \frac{1}{2}\dot q^2-\frac{m^2}{2}q^2-\frac{\alpha ^2}{2}\ddot q^2 \;\;\;\;,m>0,\;\alpha >0   \label{w4}
\end{eqnarray}
We consider mainly the oscillatory regime $2m\alpha <1$. The Euclidean version of eq.(\ref{w4}) reads
\begin{eqnarray}
L= \frac{1}{2}\dot q^2+\frac{m^2}{2}q^2+\frac{\alpha ^2}{2}\ddot q^2   \label{w5}
\end{eqnarray}
We are  interested in the propagator:
\begin{eqnarray}
K(q,\dot q.;q_0,\dot q_0;t)= \int Dq e^{-\int\limits_{0}^{t}d\tau L_E(\tau )}  \label{w6}
\end{eqnarray}
 where the integration goes over all paths obeying $q(0)=q_0, \;\dot q(0)=\dot q_0,\; q(t)=q,\; \dot q(t)=\dot q$. $K$\ is computed by standard method: first
we split $q(\tau )$\ into classical path $q_{cl}$\ obeying Lagrange equation together with the above boundary conditions and the quantum
contribution to be integrated over. The result reads
\begin{eqnarray}
K=e^{-S_E[q_{cl}]} \int D\tilde q e^{-S_E [\tilde q]}  \label{w7}
\end{eqnarray}
where
\begin{eqnarray}
S_E [\tilde q]=\int \limits_{0}^{t}d\tau \tilde q(\tau )\left(-\frac{1}{2}\frac{d^2}{d\tau ^2}+\frac{m}{2}+\frac{\alpha ^2}{2}\frac{d^4}{d\tau ^4}
\right)\tilde q(\tau )   \label{w8}
\end{eqnarray}
The integral on the RHS of eq.(\ref{w7}) can be expressed in terms of the determinant of the differential operator entering eq.(\ref{w8}). The latter
 is computed either by time discretization or by the method described in \cite{b10},   up to a normalization factor to be chosen to provide the appropriate
 Dirac delta function for $K$\ in the $t\rightarrow 0$\ limit. The final result reads
\begin{eqnarray}
&&K(q,\dot q;q_0,\dot q_0;t)=\frac{\alpha }{\pi \sqrt{2C(t)}} e^{-S_E[q_{cl}]}  \label{w9} \\
&&C(t)=\frac{2\alpha }{m} \left(\frac{sh^2(\frac{1}{2\alpha }\sqrt{1+2m\alpha }t)}{1+2m\alpha } - \frac{sh^2(\frac{1}{2\alpha }\sqrt{1-2m\alpha }t)}{1-2m\alpha }\right) \nonumber
\end{eqnarray}
Our  aim is now to construct the quantum theory (observables, the space of states, etc.) in such a way that $K$, as given by eq. (\ref{w9}), is
the propagator in the standard coordinate representation.

Naively, we could expect the following formula to hold
\begin{eqnarray}
K(q,\dot q;q_0,\dot q_0;t) =  \sum\limits_{n}\Psi  _n(q,\dot q)\overline {\Psi _n(q_0,\dot q_0)}e^{-tE_n} \label{w10}
\end{eqnarray}
where $\Psi _n \;(E_n)$\ are the eigenfunctions (eigenvalues) of the physical Hamiltonian $H$. Eq. (\ref{w10}) implies the following symmetry properties
\begin{eqnarray}
K(q,\dot q;q_0,\dot q_0;t)=\overline{K(q_0,\dot q_0;q,\dot q;t)}  \label{w11}
\end{eqnarray}
However, $K$, as given by eq.(\ref{w9}), obeys
\begin{eqnarray}
K(q,\dot q;q_0,\dot q_0;t)=\overline{K(q_0,-\dot q_0;q,-\dot q;t)}   \label{w12}
\end{eqnarray}
(actually, everything is real here).
To see this let us note that, together with any solution $q_{cl}$, $\tilde q_{cl}(\tau )\equiv q_{cl}(t-\tau )$\ is also
  a solution to classical
 equation of motion obeying the boundary conditions: 
 $\tilde q_{cl}(0)=q,\; \dot {\tilde q}_{cl}(0)=-\dot q, \;\tilde q_{cl}(t)=q_0, \;\dot {\tilde q}_{cl}(t)=-\dot q_0$;
 moreover, $S_E[q_{cl}]=S_E[\tilde q_{cl}]$.\\
 \hspace*{15pt} By comparing eqs.(\ref{w10})$\div$(\ref{w12}) one can conclude that the "physical" scalar product differs from the standard 
one ( cf. Ref. \cite{b9} and see below). \\
 The propagator $K$\ given by eq. (\ref{w9}), obeys some differential equation. To derive it we check first, using standard methods \cite{b13} 
that $S_E[q_{cl}]$\ obeys the Hamiltonian-Jacobi equation with the Ostrogradski Hamiltonian $H_E$\ derived from $L_E$,
\begin{eqnarray}
H_E=p_1q_1 + \frac{1}{2\alpha ^2}p^2_2 - \frac{1}{2}q^2_2 - \frac{m^2}{2}q^2_1 \label{w13}
\end{eqnarray}
with $q_1=q,\;q_2=\dot q$. This implies, by virtue of the identity
\begin{eqnarray}
\frac{\dot C}{C}=\frac{2}{\alpha ^2}\frac{\partial ^2S_E[q_{cl}]}{\partial q^2_2}   \label{w14}
\end{eqnarray}
(which can be verified from the explicit form of $S_E$) that $K$\ obeys
\begin{eqnarray}
\frac{\partial K}{\partial t}=\left( -q_2\frac{\partial }{\partial q_1} + \frac{1}{2 \alpha ^2}\frac{\partial ^2}{\partial q_2^2} - \frac{m^2}{2}q_1^2 - \frac{1}{2}q^2_2 \right)K    \label{w15}
\end{eqnarray}
together with the boundary condition $K(q,\dot q;q_0,\dot q_0;0)=\delta (q-q_0)\delta (\dot q-\dot q_0)$\\
In order to construct the quantum theory from our Euclidean path integral we assume, as it is the case for first-order theory, that  $K$\
is the kernel of the evolution operator $\hat K=e^{-tH}$\ in the standard coordinate representation. $\hat K$\  obeys
\begin{eqnarray}
\frac{d\hat K}{dt}=-H\hat K  \label{w16}
\end{eqnarray}
 Eq.(\ref{w15}) is then identified with the coordinate representation version of  eq.(\ref{w16}). This allows us to write the Hamiltonian
\begin{eqnarray}
H=iq_2p_1 + \frac{1}{\alpha ^2}p_2^2 + \frac{m^2}{2}q_1^2 + \frac{1}{2}q_2^2  \label{w17}
\end{eqnarray}
Unfortunately, our Hamiltonian
 is not hermitean, $H^+\not= H$. In order to cure this we introduce the metric operator $\eta $\ obeying
\begin{eqnarray}
\eta q_i\eta =(-1)^{i+1}q_i,\;\;\;\eta p_i\eta =(-1)^{i+1}p_i,\;\;\;\eta =\eta ^+,\;\;\;\eta ^2=1;
\label{w18}
\end{eqnarray}
$\eta $\ is defined by eq.(\ref{w18}) up to a sign which can be fixed by demanding $\eta |q_1,q_2 \rangle=|q_1,q_2 \rangle$.
The new ("physical") scalar product is now defined by
\begin{eqnarray}
\langle \Psi ,\Phi \rangle\equiv (\Psi ,\eta \Phi )\equiv \langle\Psi |\eta |\Phi \rangle   \label{w19}
\end{eqnarray}
One demands the observables to be hermitean with respect to the new product, i.e.
\begin{eqnarray}
 A=A^*\equiv \eta A^+\eta  \label{w20}
\end{eqnarray}
In particular $H^*=H$.
One can show that $H$, in spite of being hermitean only in generalized sense, has real positive spectrum and complete set of eigenfunctions.
This can be formally proven  by showing that one can convert $H$\ with the help of similarity transformation, into the sum of two harmonic oscillators 
\begin{eqnarray}
H=B^{-1}\left((\frac{\lambda _1^2p_1^2}{2m^2} + \frac{m^2}{2}q_1^2) + (\frac{p_2^2}{2\alpha ^2} + \frac{\alpha ^2\lambda _2^2}{2}q_2^2)\right)B \label{w21}
\end{eqnarray}
with $\lambda _{1,2}=\frac{1}{2\alpha }(\sqrt{1+2m \alpha } \mp \sqrt{1-2m \alpha })$. However $B,\;B^{-1}$\ are defined only formally
(at least one of them is unbounded  so the domains should be carefully specified and shown to contain the relevant eigenvectors
 of the harmonic
 oscillators). A more precise algebraic proof can be given  which is omitted here.\\
 So we infer the existence of complete set of eigenvectors $\Psi _{n_1,n_2}$\ obeying
\begin{eqnarray}
H \Psi _{n_1,n_2}=\left( (n_1 + \frac{1}{2})\lambda _1 + (n_2 + \frac{1}{2})\lambda _2 \right) \Psi _{n_1,n_2}  \label{w22}
\end{eqnarray}
Due to hermicity condition $H^*=H,\;\Psi _{n_1,n_2}$\ are mutually orthogonal with respect to the new scalar product; moreover, the normalization can be chosen in
such a way that
\begin{eqnarray}
\langle\Psi _{n_1,n_2}, \Psi _{n_1',n_2'} \rangle = (-1)^{n_2}\delta _{n_1, n_1'}\delta _{n_2, n_2'} \label{w23}
\end{eqnarray}
Let us expand any vector $\Phi $\ in terms of basic vectors $\Psi _{n_1,n_2}$\
\begin{eqnarray}
\Phi = \sum\limits_{n_1,n_2=0}^{\infty } \alpha _{n_1,n_2} \Psi _{n_1,n_2}  \label{w24}
\end{eqnarray}
Then, by virtue of eq.(\ref{w23}),
\begin{eqnarray}
\alpha _{n_1,n_2} = (-1)^{n_2} \langle \Psi _{n_1,n_2}, \Phi \rangle \equiv  (-1)^{n_2} \left(\Psi _{n_1,n_2}, \eta \Phi \right)  \label{w25}
\end{eqnarray}
and
\begin{eqnarray}
\Phi = \sum\limits_{n_1,n_2=0}^{\infty }(-1)^{n_2} \left(\Psi _{n_1,n_2}, \eta \Phi \right)  \Psi _{n_1,n_2}  \label{w26}
\end{eqnarray}
One can easily compute the new scalar product in coordinate representation:
\begin{eqnarray}
\langle \Phi , \Psi \rangle = \int dq_1dq_2 \langle \Phi |\eta |q_1,q_2 \rangle \langle q_1, q_2|\Psi \rangle = \int dq_1dq_2 \overline{\Phi (q_1,-q_2)}\Psi (q_1,q_2);     \label{w27}
\end{eqnarray}
Let us now define the kernel $K(q_1,q_2;q_{01},q_{02};t)$\ by the equation
\begin{eqnarray}
\langle\Phi _2, e^{-tH}\Phi _1 \rangle \equiv  \int dq_1dq_2dq_{01}dq_{02} \overline{\Phi _2(q_1,-q_2)}K(q_1,q_2;q_{01}q_{02};t)\Phi _1(q_{01},q_{02})   \label{w28}
\end{eqnarray}
Then
\begin{eqnarray}
&&\langle\Phi _2, e^{-tH}\Phi _1 \rangle = \left(\Phi _2, \eta e^{-tH}\Phi _1 \right) = \left(\eta \Phi _2,  e^{-tH}\Phi _1 \right) =      \label{w29} \\
&&\sum\limits_{n_1,n_2}(-1)^{n_2} \left(\eta \Phi _2, \Psi _{n_1,n_2} \right)  e^{-tE_{n_1,n_2}} \left(\Psi _{n_1,n_2}, \eta \Phi _1 \right)= \nonumber \\
&&\sum\limits_{n_1,n_2}(-1)^{n_2} \left(\eta \Phi _2, \Psi _{n_1,n_2} \right)  e^{-tE_{n_1,n_2}} \left(\eta  \Psi _{n_1,n_2},  \Phi _1 \right) \nonumber
\end{eqnarray}
or
\begin{eqnarray}
K(q_1,q_2;q_{01},q_{02};t) = \sum\limits_{n_1,n_2}(-1)^{n_2} e^{-tE_{n_1,n_2}}\Psi _{n_1,n_2}(q_1,q_2)\overline{ \Psi _{n_1,n_2}(q_{01},-q_{02})} \label{w30}
\end{eqnarray}
Now, $K$\ given as above, obeys  eq.(\ref{w15}) and the appropriate boundary condition \\
$K(q,\dot q;q_0,\dot q_0;0)=\delta (q-q_0)\delta (\dot q-\dot q_0)$. Therefore, we conclude that 
the Euclidean path
integral represents the evolution operator for the theory defined by the Hamiltonian(\ref{w17}) and the scalar product (\ref{w19}).\\
It is worth to notice that our conclusions concerning nonhermicity of the Hamiltonian and the stucture of scalar product agree with
 those contained in Ref. \cite{b9}.\\
 
Let us discuss the above results from the point of view presented in the Introduction. To this end consider the classical dynamics 
generated by the  Hamiltonian (\ref{w17}). It reads
\begin{eqnarray}
&& \dot q_1 = iq_2, \;\;\;\;\;\;\;\;\;\dot q_2 = \frac{1}{\alpha ^2}p_2  \label{w31} \\
&& \dot p_1 = -m^2q_1,\;\;\;\; \dot p_2 = -q_2 - ip_1  \nonumber
\end{eqnarray}
and results in the following equation for the variable $q\equiv q_1$
\begin{eqnarray}
\alpha ^2q_1^{(IV)} + \ddot q_1 + m^2q_1 = 0  \label{w32}
\end{eqnarray}
which is the Lagrange equation we have started with. Moreover, by inspecting eq. (\ref{w31}) we find that it is consistent to assume that $q_1$\ and $p_1$\
are real while $q_2$\ and $p_2$\ are purely imaginary. The variables $q=q_1$, $\dot q = iq_2$, $\ddot q = \frac{i}{\alpha ^2}p_2$\ and 
$\dddot q = \frac{-i}{\alpha ^2}q_2 + \frac{1}{\alpha ^2}$, characterizing uniquely the classical state, are real. Thus eq. (\ref{w31}) present
 the consistent   immersion of the initial  dynamics in the extended, complex one; only the auxiliary dynamical variables take complex values.\\
 At the quantum level the expectation values of $q_1$\ and $p_1$\ are real; for $q_2$\ (as well as $p_2$), on the contrary, one gets
\begin{eqnarray}
\overline{\langle \Phi , q_2 \Phi \rangle} = \langle q_2 \Phi ,  \Phi \rangle = \langle \Phi , q_2^* \Phi \rangle = - \langle \Phi , q_2 \Phi \rangle  \label{w33}
\end{eqnarray}
because $q_2^* = - q_2$.\\
Let us note that the propagator can be also formally obtained from the Hamiltonian form of path integral provided the Hamiltonian (\ref{w17}) is used
(cf. also Ref. \cite{b10} ):
\begin{eqnarray}
e^{-tH} = \int Dq_1Dq_2Dp_1Dp_2e^ {\int\limits_0^td\tau \left(i(p_1\dot q_1+p_2\dot q_2) - H(\tau ) \right)}  \label{w34}
\end{eqnarray}
Another consistency check is provided by taking the $t\rightarrow \infty $\ limit of the propagator (\ref{w9}). By virtue of eq. (\ref{w30}) one finds
\begin{eqnarray}
K(q_1,q_2;q'_1,q'_2) \sim e^{-\frac{1}{2}(\lambda _1 + \lambda _2)t}\Psi _{0,0}(q_1,q_2)\overline{\Psi _{0,0}(q'_1,-q'_2)}  \label{w35}
\end{eqnarray}
which can be easily verified to hold true.\\

Few remarks concerning the Heisenberg picture are in order. Heisenberg equations take the same form as the classical Hamilton equation (\ref{w31}).
 They preserve the $\ast $-hermitean (or $\ast $-antihermitean) character of dynamical variables; on the other hand the $+$-hermicity of canonical 
 variables is not preserved: in general $ q_i^+(t) \not= q_i(t), p_i^+(t) \not= p_i(t)$ even if $q_i^+(0) = q_i(0), p_i^+(0) = p_i(0)$. However, 
this poses  no problem since the standard Hilbert space structure plays only auxiliary role in our considerations; we could have started from the
indefinite metric space from the very beginning.\\

Up to now we were considering the oscillatory regime $2m \alpha < 1 $. The natural question arises whether one obtains a reasonable quantum theory for $2m \alpha  \geq  1$. By a closer inspection of the way eq.(\ref{w9}) has been derived we immediately conclude that it is valid also in the region  $2m \alpha  \geq  1$. However, the preexponential factor in eq.(\ref{w9}) gets modified: 
\begin{eqnarray}
C(t)=\left\{\begin{array}{cl}
\frac{2\alpha }{m} \left(\frac{sh^2(\frac{1}{2\alpha }\sqrt{1+2m\alpha }t)}{1+2m\alpha } - \frac{sin^2(\frac{1}{2\alpha }\sqrt{2m\alpha -1}t)}{2m\alpha-1 }\right) & for \;\; 2m\alpha > 1     \\
2\alpha ^2sh^2(\frac{t}{\sqrt2\alpha }) - t^2  & for \;\; 2m = 1   
\end{array}\right.    \label{w36}
\end{eqnarray} 
Eq.(\ref{w9}) shows that for $2m\alpha  > 1$\ the exponential damping is now modulated by oscillating factors. Comparing this with the (assumed)
 form of $K$\, eq. (\ref{w30}), one concludes that the energies cannot be purely real. This conclusion is confirmed by the direct analysis 
of spectral properties of the Hamiltonian (\ref{w17}). It can be shown that for $2m \alpha  >  1$\ it posses the purely point complex spectrum
\begin{eqnarray} 
E_{n_1,n_2} = \frac{n_1 + n_2}{2\alpha } \sqrt{1 + 2m\alpha } + i\frac{n_1-n_2}{2\alpha }\sqrt{2m\alpha  - 1} ;  \label{w37}
\end{eqnarray}
note that $\overline{E_{n_1,n_2}} = E_{n_1,n_2}$\ and as a result of reality of eigenequation, $\overline{\Psi _{n_1,n_2}} = \Psi _{n_2,n_1}$;
moreover, ${\Psi _{n_1,n_2}}$\ form a (nonorthogonal) basis in the space of states; \\
For $2m\alpha  = 1$ the exponential damping is modified by the powers of $t$. Again, this agrees with the spectral properties of $H$: 
for $2m\alpha  = 1$ its spectrum consists of the eigenvalues $E_n = \frac{n}{\sqrt{2}\alpha }$ and  to any $n$\ there corresponds an
($n+1$)-dimensional invariant subspace where $H$\ takes the Jordan block form
(cf. \cite{b14}  for other examples of nondiagonalizable Hamiltonians).\\

We conclude that the Euclidean path integral gives the proper expression for the kernel of $e^{-tH}$\ also for $2m\alpha  \geq  1$.
In particular, the reality of $K$\ follows directly from the properties of the eigenvalues and eigenfunctions under complex conjugation.
However, the resulting theory posses complex energy spectrum.\\
\section{Field theory}
Let us now consider the field-theoretic analogy of PU oscillator. We start with the Lagrangian
\begin{eqnarray}
&&\mathcal {L} = \frac{1}{2} \Phi ( \square + m_1^2)(\square +m_2^2)\Phi =  \frac{1}{2} (\square\Phi)^2 + 
\frac {m_1^2 + m_2^2}{2}\Phi \square \Phi + \label{w38} \\
&&\;\;\;\;\;\frac{1}{2} m_1^2m_2^2 \Phi ^2 + b.t., \;\;\;\;\; m_1^2 > m_2^2 \nonumber 
\end{eqnarray}
In analogy with eqs.( \ref{w31}) we put 
\begin{eqnarray}
\Phi = \Psi _1\;\;\;\;\; \dot \Phi = i\Psi _2  = \dot \Psi _1 \;\;\;\;\; \ddot\Phi = i\dot\Psi _2 \label{w39}
\end{eqnarray}
and define  the extended Lagrangian
\begin{eqnarray}
&&\mathcal{\tilde {L}}(\Psi ,\dot \Psi ,\Psi _2,\dot \Psi _2,\lambda ) \equiv  \label{w40}\\
&& \frac{1}{2} (i\dot \Psi _2 - \triangle \Psi _1)^2 + \frac {m_1^2 + m_2^2}{2}\Psi _1( i\dot \Psi _2 - \triangle \Psi _1) + \frac{1}{2} m_1^2m_2^2 \Psi _1^2 + \lambda (\dot \Psi _1 - i\Psi _2)  \nonumber 
\end{eqnarray}
The theory is singular due to the appearance of Lagrangian multipliers $\lambda $. Let
\begin{eqnarray}
\tilde L \equiv \int d^3 x \mathcal{\tilde {L}} \label{w41}
\end{eqnarray}
The canonical momenta read
\begin{eqnarray}
&&\pi = \frac {\delta \tilde L}{\delta \dot \lambda } =  0 \nonumber \\
&&p_1 = \frac {\delta \tilde L}{\delta \dot \Psi _1 } = \lambda  \\ \label{w42}
&&p_2 = \frac {\delta \tilde L}{\delta \dot \Psi _2 } = -\dot \Psi _2 +i\left(- \triangle \Psi _1 +\frac{i}{2}(m_1^2 + m_2^2)\right)\Psi _1 \nonumber
\end{eqnarray}
so there are two sets of primary constraints
\begin{eqnarray}
&&\pi \approx  0 \nonumber \\
&&p_1 - \lambda  \approx  0  \label{w43}
\end{eqnarray}
while the Hamiltonian density reads 
\begin{eqnarray}
&&\mathcal {H} = -\frac{1}{2} p_2^2 - ip_2\triangle \Psi _1 + \frac{i}{2} (m_1^2 + m_2^2) p_2\Psi _1 + ip_1\Psi _2 - 
\frac{1}{2}m_1^2m_2^2\Psi _1^2 \nonumber \\ 
&&\;\;\;\;\;\;+ \frac{1}{8} (m_1^2 + m_2^2)\Psi _1^2   + \rho \pi _\lambda  + \sigma (p_1 - \lambda )     \label{w44}
\end{eqnarray}
yielding the canonical equations
\begin{eqnarray}
&&\dot \lambda  = \rho  \nonumber \\
&&\dot \pi  = \sigma  \nonumber \\
&&\dot \Psi _1 = i\Psi _2 + \sigma  \label{w45} \\
&&\dot \Psi _2 =  -p_2 - i\triangle \Psi _1 + \frac{i}{2} (m_1^2 + m_2^2) \Psi _1 \nonumber \\
&&\dot p_1 = i\triangle p_2 - \frac{i}{2} (m_1^2 + m_2^2) p_2 - \frac{1}{4} (m_1^2 + m_2^2)^2\Psi _1 + m_1^2m_2^2\Psi _1 \nonumber \\
&&\dot p_2 = -ip_1   \nonumber
\end{eqnarray}
Eqs.(\ref{w43}) and (\ref {w45}) yield
\begin{eqnarray}
&&\sigma =  0 \nonumber \\
&&\rho = i\triangle p_2 - \frac{i}{2} (m_1^2 + m_2^2) p_2 - \frac{1}{4} (m_1^2 + m_2^2)^2\Psi _1 + m_1^2m_2^2\Psi _1  \label{w46}
\end{eqnarray}
Eqs.(\ref{w43}) are the only constraints;  they are of second kind and allow to eliminate $\lambda $\
 and $\pi $. The Dirac brackets for remaining variables take the standard form.\\
 It is now straightforward to construct the quantum counterpart of our theory. We skip standard details
  and write out the final result:
 \begin{eqnarray}
&& \Phi  _1(x) = \sum\limits_{i=1}^{2} \int d^3k \left( a_i(k)e^{-ik_ix} +  {a_i^\ast (k)}e^{ik_ix} \right)  \label{w47} \\
&& k_i = \left( \sqrt {\vec k^2 + m_i^2}, \vec k \right)  \nonumber 
\end{eqnarray}
where
 \begin{eqnarray}
[a_i(\vec k), a_k^\ast (\vec k')] = \frac{(-1)^{i+1} \delta _{ik}}{(m_1^2 - m_2^2)k_i^0}\delta ^{(3)}(\vec k - \vec k') \label{w48}
\end{eqnarray}
Redefining
 \begin{eqnarray}
a_i(\vec k) = \frac{b_i(\vec k_i)}{ \sqrt{(m_1^2 - m_2^2)k_i^0}} \label{w49}
\end{eqnarray}
we arrive at the (almost) standard commutation rules
  \begin{eqnarray}
[b_i(\vec k), b_k^\ast (\vec k')] = (-1)^{i+1} \delta _{ik} \delta ^{(3)}(\vec k - \vec k') \label{w50}
\end{eqnarray}
The space of states is spanned by the fourmomentum eigenvectors describing particles of two types created by the operators $a_i(\vec k)$.
The subspace of the positive norm consists of states carrying an even number of particles of type $"2"$.\\
Let us consider the energy-momentum tensor. It is not difficult to find its symmetriezed  form
 \begin{eqnarray}
&&T^{\mu \nu } = 2\partial ^\mu \partial ^\nu \Phi \square \Phi - \partial ^\mu \Phi\square \partial ^\nu \Phi - \partial ^\nu  \Phi\square 
\partial ^\mu  \Phi  + \nonumber \\
&&\;\;\;\;\;\;\partial ^\mu \partial ^\nu \partial _\alpha \Phi \partial ^\alpha \Phi  - \partial ^\mu \partial _\alpha \Phi \partial ^\nu \partial ^\alpha \Phi - \frac{1}{2} g^{\mu \nu }(\square\Phi )^2 + \nonumber \\
&&\;\;\;\;\;\; \frac{1}{2}(m_1^2 + m_2^2)(\Phi \partial^ \mu \partial ^ \nu \Phi  - \partial^\mu \Phi \partial ^\nu \Phi  - g^{\mu \nu }\Phi \square \Phi ) - \nonumber \\
&& \;\;\;\;\;\;\frac{1}{2} g^{\mu \nu }m_1^2 m_2^2\Phi ^2  \label{w51}
\end{eqnarray}
One easily checks that, in spite of its form, $T^{\mu \nu }$\ depends on time derivatives of $\Phi $\ up to third order only. In particular, expressing
$\Phi  = \Psi _1, \dot\Phi , \ddot\Phi $\ and $ \dddot \Phi $\ in terms of canonical variables (with the help of eqs.(\ref{w45}) )
we find that $T^{00} $\ coincides, up to some spatial boundary terms, with the Hamiltonian density (\ref{w44}). In terms of $\Phi $\ and its
derivatives it reads
\begin{eqnarray}
&&T^{00} = \frac{1}{2} \ddot \Phi ^2 - \dot \Phi \dddot \Phi  - \frac{1}{2}(\bigtriangleup \Phi )^2 + \dot \Phi \bigtriangleup \dot \Phi  + \nonumber \\
&&\;\;\;\;\;\;\frac{1}{2} (m_1^2 + m_2^2) (\Phi \bigtriangleup \Phi  - \dot \Phi ^2) - \frac{1}{2} m_1^2 m_2^2\Phi ^2  \label{w52} 
\end{eqnarray}
Inserting here the general solution (\ref{w47}) one gets 
\begin{eqnarray}
\int d^3x T^{00} = \int d^3 \vec k \left(k_1^0b_1^\ast (\vec k)b_1(\vec k) + k_2^0b_2^\ast (\vec k)b_2(\vec k )\right) \label{w53} 
\end{eqnarray}
The energy has positive eigenvalues and conserves the particle numbers. The latter holds true also for other generators of Poincare group.
Therefore, the subspace of positive definite scalar product is invariant under the action of Poincare group.
 The theory is perfectly consistent.

\section{Conclusions}

Let us compare the Hawking-Hertog approach with the standard one based on Ostrogradski formalism. In both cases the dynamics of the original
variable $q$\ is immersed in a wider one. In the standard case all additional variables are kept real while within the approach  inspired by
 the Euclidean path integral some of them attain complex values. This difference results in difference in the form of the Hamiltonian. 
 The Ostrogradski Hamiltonian is hermitean but unbounded from below; on the other hand, the Hamiltonian following 
from Euclidean path integral is nonhermitean which implies the necessity of modifying the definition of scalar product.
In the oscillatory regime $2m\alpha <1$\ the nonhermitean Hamiltonian has real spectrum which, in contrast to that of Ostrogradski one,
is bounded from below. For $2m\alpha \geq 1$\ the Ostrogradski Hamiltonian continues to be hermitean; its spectrum, however,
 becomes continuous (and still unbounded from below). On the other hand the  Hamiltonian (\ref{w17}) has a purely point spectrum 
even for $2m\alpha  > 1$\ (while it is not diagonalizable if $2m\alpha  = 1$) but the eigenvalues are no longer real (although the real 
parts are positive). Therefore, it cannot be considered as a viable quantum mechanical Hamiltonian; its defects cannot be cured by any 
change of metrics in the space of states. It must be, however, stressed that the Euclidean path integral still represents the operator
$e^{-tH}$.\\

Having a detailed knowledge of the structure of the space of states one can specify the kind of perturbations which generate the
 unitary evolution in the subspace corresponding to the eigenvalue $1$\ of metric operator $\eta $. They should be represented
by the  operators hermitean 
both in the usual as well as generalized sense. Then the evolution operator commutes with $\eta $\ and leaves invariant the above 
subspace. We should, however, assume that the interaction Lagrangian does not contain second derivatives; if this is not 
the case the theory will be probably not well defined.\\

Similar results hold true for the field-theoretical counterpart of PU oscillator.
We conclude that Hawking-Hertog method provides, at least in some specific cases, a reasonable, consistent method of quantization
for theories containing second derivatives quadratically.\\

The necessary condition for the applicability of H-H method is the positive definiteness of Euclidean action. However, as we have seen,
this by far not sufficient; the Euclidean action for PU oscillator continues to be positive definite for $2m\alpha \geq 1$\ while 
the resulting Hamiltonian has complex eigenvalues or is not diagonalzable.

\vspace*{15pt}
{\large\bf Acknowledgement}

We are grateful to Piotr Kosi\'nski and Maciej Przanowski for helpful discussions.


\begin{thebibliography}{99}
\bibitem{b1}
K. S. Stelle , Phys. Rev.  \underline{D16} (1977), 953. \\
E. S. Fradkin  , Tseytlin A.A., Nucl. Phys. \underline{B201}, (1982), 469.
\bibitem{b2} 
A. M. Polyakov ,Nucl. Phys. \underline{B268}, (1986), 406.
\bibitem{b3}
M. S. Plyushchay ,  Mod. Phys. Lett.\underline{A3}, (1988), 1299;
 Nucl. Phys. \underline{B362}, (1991), 54; Phys. Lett.\underline{B262} (1991), 72 ; Mod. Phys. Lett.\underline{A4}  (1989), 837;
 Phys. Lett.\underline{B243}, (1990), 383 ; Electron. Journ. Theor. Phys. \underline{3N10}, (2006), 17.  \\
\bibitem{b4}
M. R. Douglas , N. A. Nekrasov , Rev. Mod. Phys.\underline{73}  (2001), 997.\\
R. J. Szabo , Phys. Rep.\underline{378}, (2003), 207.
\bibitem{b5}
A. Mironov, A. Morozov, Theor. Math. Phys. \underline{156}  (2008), 1209;  Int. J. Mod. Phys. \underline{A23}  (2008), 4686. \\
D. Galakhov, JETP Letters \underline{87}, (2008), 452.
\bibitem{b6}
M. Ostrogradski, Mem. Ac. St. Petersburg \underline{VI 4}, (1850), 385.\\
J. Govaerts, M. S. Rashid,  arXiv:hep-th/9403009.\\
T. Nakamura,   S. Hamamoto,  Prog. Theor. Phys. \underline{95}, (1996), 409.
\bibitem{b7}
S. W. Hawking  "Quantum Field Theory and Quantum Statisics: Essays in honour of the 60th Birthday of E.S.Fradkin"
eds.  C. J. Batalin,  C. A. Isham, C. A. Vilkovisky, Hilger, Bristol, (1987).
\bibitem{b8}
A. Pais, G. E. Uhlenbeck , Phys. Rev.\underline{79}, (1950), 145.
\bibitem{b9}
D. Boulware, D. Gross, Nucl. Phys. \underline{B233}, (1984), 1.
\bibitem{b10}
H. Kleinert, Journ. Math. Phys.\underline{27}, (1986), 2003
\bibitem{b11}
S. Hawking, T. Hertog, Phys. Rev. \underline{D65}, (2002), 103515.
\bibitem{b12}
T. Dreyfus, H. Dym, Duke Math. Journ.\underline{45}, (1978),15.
\bibitem{b13}
L. Landau, E. Lifshitz, " Mechanics", Pergamon Press, (1968).
\bibitem{b14}
W.D. Heiss, Journ. Phys. \underline{A37}, (2004), 2455.
\end{thebibliography}
\end{document}